\documentclass[aps,prd]{revtex4}
\usepackage{subfigure,graphics} 
\usepackage{flafter} 
\begin{document} 

\title{
Dissipation coefficients for supersymmetric inflatonary models} 
\author{Ian G. Moss}
\email{ian.moss@ncl.ac.uk}
\author{Chun Xiong}
\affiliation{School of Mathematics and Statistics, University of  
Newcastle Upon Tyne, NE1 7RU, UK}

\date{\today}


\begin{abstract}
Dissipative effects can lead to a friction term in the equation of motion for
an inflaton field during the inflationary era. The friction term may be linear
and localised, in which case it is described by a dissipation coefficient.
The dissipation coefficient is calculated here in a supersymmetric model with a
two stage decay process in which the inflaton decays into a thermal gas of
light particles through a heavy intermediate. At low temperatures, the
dissipation coefficient $\propto T^3$ in a thermal approximation. Results are
also given for a non-equilibrium anzatz. The dissipation coefficient is
consistent with a warm inflationary regime for moderate ($\sim 0.1$) values of
the coupling constants.
\end{abstract}
\pacs{PACS number(s): }

\maketitle
\section{introduction}
Inflationary models have made a huge contribution to our understanding of the
early universe and the origin of the basic cosmological features which are
observed today in ever increasing detail \cite{guth81,linde82,albrecht82}. 
This paper is concerned with the
mechanism for converting the vacuum energy which dominates the
universe during inflation into the radiation which dominates after inflation.
Radiation can be produced during inflation, at the end of inflation or after
inflation has finished. We will consider the first possibility in a particular
class of supersymmetric models, to see whether a significant amount of
radiation can be production during inflation. When this happens, the situation
is described as warm inflation \cite{bererafang95,berera95} (see also
\cite{moss85}).

The production of radiation is balanced by
dissipative terms in the equation of motion of the homogeneous inflaton field
$\phi$. 
The simplest possibility is a linear dissipation term,
\begin{equation}
\ddot\phi+(3H+\Gamma)\dot\phi+V_\phi=0
\end{equation}
where $V_\phi$ is the $\phi$ derivative of the inflaton potential, $H$ is the
expansion rate of the universe and $\Gamma$, which in general will depend on
both $\phi$ and temperature $T$,  is called the dissipation coefficient. The
dissipation coefficient was originally introduced as an arbitrary parameter
into discussions of reheating
\cite{Albrecht:1982mp}. Explicit formulae for the dissipation coefficient were
obtained from analysing particle production
\cite{Abbott:1982hn,Morikawa:1984dz} and linear response theory
\cite{hosoya84}. Later work has tended
to use the Schwinger-Keldeysh approach which we use here, eg 
\cite{berera96,berera97,berera98,moss02,berera01,berera03,lawrie01,Lawrie:2002zd}.

Dissipation can lead to important consequences when $T>H$ (called the weak
regime of warm inflation) and when $\Gamma>H$ (the strong regime). In the weak
regime, the primordial density perturbation spectrum is determined by thermal
fluctuations rather than vacuum fluctuations. In the strong regime,
restrictions on the gradient of the inflaton potential may be relaxed
\cite{berera95}. This is especially important in the context of $F$-term
supersymmetric inflationary models  (see \cite{lyth98} for a review). The
potentials in these models contain supergravity corrections which rule out the
ordinary type of inflation, but not warm inflation
\cite{berera03,Hall:2004zr}.

On the other hand, the calculations which have been done so far have assumed
that radiation remains close to thermal equilibrium, and this leads to
additional consistency requirements:
\begin{itemize}
\item Thermal corrections to the inflaton potential have to be consistent with
the restrictions on the gradient of the potential.
\item The thermalisation time $\tau$ of the radiation should be smaller than
the expansion timescale of the universe.
\end{itemize}
These conditions lead to strong restrictions on the existence of warm
inflation. Nevertheless, it is useful to investigate the close to
equilibrium regime before moving on to the more difficult far from equilibrium
regime.

The transfer of energy from the inflaton to another field $\chi$ is
dependent the coupling strength of the interaction, the relative 
sizes of the mass of the inflaton $m_\phi$, the mass of decay product $m_\chi$
and the  temperature $T$. The theory of reheating is well-developed for the
post-inflationary era when the inflaton is oscillating and the parameter range
$m_\phi>m_\chi$. According
to linear theory, the rate of decay of these oscillations is determined by the
decay rate for processes such as $\phi\to 2\chi$ 
\cite{Abbott:1982hn,Traschen:1990sw,Yokoyama:2005dv,Yokoyama:2006wt}.
The theory of preheating is the modification which takes into account
non-linear effects close to the particle production thresholds
\cite{kofman94,Boyanovsky:1994me,Boyanovsky:1995em}.

The theory of the dissipation coefficient in the inflationary era is also
quite well-developed, especially at high temperature, when $T>m_\chi$. This is
the case in which the consistency 
requirements for an inflationary model are most severe, with large thermal 
corrections when the radiation is strongly coupled and large relaxation 
times when the radiation is weakly coupled
\cite{yokoyama99}.

Recently, there has been interest in the possibility that the inflaton
can dissipate energy at low temperature via a two-stage decay 
$\phi\to\chi\to yy$, with the heavy field $\chi$ acting
as  an intermediary in producing a light field $y$
\cite{Berera:2001gs,berera02,Berera:2004kc}. The dissipation
in these models can become a significant feature if the coupling strengths of
the inflaton to the heavy field is moderately large.

Inflation is usually associated with a very weakly coupled inflaton to protect 
the gradient of the potential from picking up large quantum corrections.
However, in supersymmetric models the restrictions on the couplings are less
severe. The gradient of the inflaton potential can be small enough to allow
inflation whilst the thermalisation time for the radiation remains short,
allowing warm inflation for moderate values of the coupling constants in
these models \cite{Bastero-Gil:2004tg,Hall:2004zr,Bastero-Gil:2005gy}.

In this paper, we shall calculate the dissipation coefficients in
supersymmetric models which have an inflaton together with multiplets of heavy
and light fields. Different contributions to the final result can be applied
also to non-supersymmetric models.

In the high temperature regime $T>m_\chi$, our results are similar in many ways
to those obtained previously for self-interacting \cite{hosoya84} 
and coupled systems
\cite{gleiser94}.  There
are some differences for $T\sim m_\chi$ due to the details of the $\chi\to yy$
decay. These differences affect the temperature dependence of the dissipation
coefficient. This could potentially have observable consequences, since studies 
of density perturbations in models with different forms 
of dissipation terms have shown that the spectrum of density perturbations is 
quite sensitive to the temperature dependence of the dissipation coefficient
\cite{Hall:2003zp}.

We have new results for the regime  $H<T<m_\chi$ and $m_\phi<m_\chi$. Previous
calculations in this regime have been, in effect, highly non-thermal
\cite{Morikawa:1984dz,Berera:2001gs}. 
Our results differ substantially from these non-thermal calculations.
We find that the leading order dissipation terms in the inflaton equation
vanish 
in the low temperature limit, but they are sufficiently large to ensure that
the
temperature does not fall to zero during inflation. We shall see in section VI 
that warm inflation appears to be allowed for moderate values of the coupling
constants.

\section{Thermal field theory}

We begin with some of the basic notions and results from thermal field theory
which will be used later on in this paper.  More details and proofs can be
found in standard texts, eg \cite{lebellac}.

The ensemble
average of a Heisenberg operator $\hat A$ can be written
\begin{equation}
\langle \hat A(x) \rangle=
{\rho_{ii'}\langle\psi_i|\hat A |\psi_{i'}\rangle\over
\rho_{ii'}\langle\psi_i|\psi_{i'}\rangle}
={{\rm tr}(\rho\hat A)\over {\rm tr}(\rho)}
\end{equation}
where $\rho$ is the density matrix and repeated indices indicate sums.
We shall be using a shifted version of the Schwinger-Keldysh, or closed-time
path formalism to evaluate ensemble averages \cite{schwinger61,keldysh64}. 
For a scalar field $\phi$, the generating functional
\begin{equation}
Z[\phi_1,\phi_2,J_1,J_2]=
\rho_{ii'}\langle\psi_{i}|
\,T^*\exp\left(-i\int J_2(\hat\phi-\phi_2\,)\,\right)
T\exp\left(i\int
J_1(\hat\phi-\phi_1)\,\right)|\psi_{i'}\rangle,
\end{equation}
where $T$ denotes time ordering of the operators with the smallest time on
the right and $T^*$ denotes time ordering of the operators with the smallest
time on the left. Shifting the field operators by $\phi_1$ and $\phi_2$ is done
for later convenience.

Following Calzetta and Hu \cite{calzetta88}, we combine the sources into a pair
$J_a=(J_1,J_2)$ and introduce a metric $c_{ab}={\rm diag}(1,-1)$ to raise and
lower the index $a$.  This removes the inconvenience with the relative signs of
the two source terms. The generating functional gives rise to four two-point
functions,
\begin{equation}
G_{ab}=-\left.{i\over Z}
{\delta^2 Z\over \delta J^a\delta J^b}\right|_{(J_1,J_2)=0}
\end{equation}
If we let $\hat\eta=\hat\phi-\phi_a$, these two-point functions can be
expressed in terms of the functions
\begin{equation}
G^>(x,x')=G^<(x',x)=i\langle \hat\eta(x)\hat\eta(x')\rangle.\\
\end{equation}
The individual components are
\begin{eqnarray}
G_{11}(x,x')&=&\theta(t-t')\,G^>(x,x')\,\,+\theta(t'-t)\,G^<(x,x')\label{gee}\\
G_{22}(x,x')&=&\theta(t-t')\,G^<(x,x')+\theta(t'-t)\,G^>(x,x')\\
G_{12}(x,x')&=&G^<(x,x')\\
G_{21}(x,x')&=&G^>(x,x')\label{geh}
\end{eqnarray}
The two-point function $G_{11}$ is the thermal analogue of the Feynman green
function $G_F$. (Placing a factor of $i$ in front
of the expectation values in the definition of the two-point functions follows
\cite{weinberg} but differs from most references.)

Consider a system which remains close to local equilibrium. 
The natural starting point 
for perturbation theory is  a free theory with an initially thermal
distribution of states. The two-point functions for the interacting theory are
given by Feynman diagram expansions with lines representing free theory
propagators $-iG_{0ab}$ and vertices derived from an interaction Lagrangian.
The mass of the propagators and the vertices depend on the background fields
$\phi_a$, which may be time or space dependent. Each of these vertices has a
label $a$ attached, and the final expression for the diagram has a sum over
the $a$'s. Later, we will make use of the following result,
\begin{itemize}
\item
The maximum time rule. The contribution from a diagram vanishes if the time at
any vertex is larger than all of the times on the external lines of the
diagram.
\end{itemize}
This follows directly from the operator form of the two-point functions
(\ref{gee}-\ref{geh}).

The next step is to consider the background fields $\phi_a$. These can be
identified with the ensemble averages
of the field operator by setting
\begin{equation}
{\delta Z\over\delta J^a}=0.
\end{equation}
We assume that this equation can be solved for $J_a[\phi_b]$. The
effective field equation is then $F=0$, where
\begin{equation}
F[\phi](x)=-\left.{\delta \Gamma[\phi_a]\over\delta\phi_b(x)}
\right|_{\phi_a=\phi}=0,
\end{equation}
and the effective action
\begin{equation}
\Gamma[\phi_a]=-i\ln Z[\phi_a,J_b[\phi_c]].
\end{equation}
The effective action can be expanded, as usual, by a sum over 1-particle
irreducible vacuum diagrams, with internal lines representing the free-theory
propagators $-iG_{0ab}$.

Dirac and Majorana fields can be included in the Schwinger-Keldysh formalism
without any difficulty. We define the fermion two-point functions
\begin{equation}
S^>(x,x')=i\langle\psi(x)\overline\psi(x')\rangle,\quad
S^<(x,x')=-i\langle\overline\psi(x')\psi(x)\rangle\label{fprop}
\end{equation}
and
\begin{eqnarray}
S_{11}(x,x')&=&\theta(t-t')\,S^>(x,x')\,\,+\theta(t'-t)\,S^<(x,x')\\
S_{22}(x,x')&=&\theta(t-t')\,S^<(x,x')+\theta(t'-t)\,S^>(x,x')\\
S_{12}(x,x')&=&S^<(x,x')\\
S_{21}(x,x')&=&S^>(x,x')
\end{eqnarray}
where $\overline S^>(x,x')=- S^>(x',x)=\widetilde S^<(x,x')$. (For a Dirac
matrix $M$, let $\overline M=\gamma^0 M^\dagger\gamma^0$ and 
$\widetilde M=(C^{-1}MC)^T$, where $C$ is the charge conjugation matrix; 
$\gamma^\mu=\overline\gamma^\mu=-\widetilde\gamma^\mu$.)

\begin{center}
\begin{figure}[ht]
\scalebox{0.5}{\includegraphics{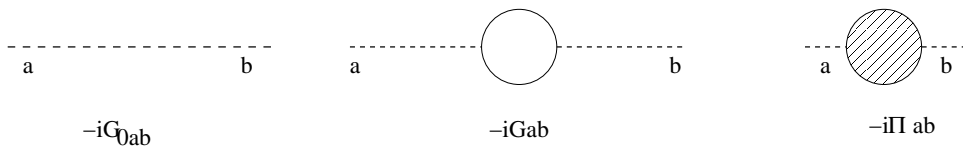}}
\caption{Diagrammatic representation for the free two-point function
$-iG_{0ab}$, the two-point function $-iG_{ab}$ and
the vacuum energy $-i\Pi_{ab}$.
}
\label{diagrams}
\end{figure}
\end{center}

Finally, we consider an interacting system in thermal equilibrium with
temperature $T$. The background fields must then be constant, $\phi_1=\phi_2$
and the scalar two-point functions 
$G_{ab}(x,x')\equiv G_{ab}(x-x')$. We set
\begin{equation}
G_{ab}(P)=\int d^4x\,G_{ab}(x,x')\,e^{-iP\cdot(x-x')}
\end{equation}
where the 4-momentum $P=({\bf p},\omega)$. The vacuum polarization or
self-energy $\Pi_{ab}$ is defined by
\begin{equation}
G^{-1}{}^{ab}(P)=(P^2+m^2)c^{ab}+\Pi^{ab}.
\end{equation}
A diagrammatic representation for the two-point functions is shown in figure
\ref{diagrams}.
We shall often represent the dissipative part of the self-energy by a function
$\alpha$, defined by
\begin{equation}
\alpha=i(\Pi_{21}-\Pi_{12})\label{adef}
\end{equation}
The function $\alpha$ is related to the relaxation time $\tau_p$ of the
particle state via the relation $\alpha=2\omega_p\tau_p^{-1}$, where
$\omega_p$ is the energy of a state with momentum $p$.

In the case of fermions, we denote the self-energy  by $\Sigma_{ab}$, and
\begin{equation}
S^{-1}{}^{ab}(P)=(\gamma\cdot P+m)c^{ab}+\Sigma^{ab}.
\end{equation}
Further properties of the equilibrium propagators are listed in the appendix.

\section{dissipation terms}

\subsection{Adiabatic approximations}

We shall examine the effective field equation for an inflaton field which is
interacting with radiation and evolving very slowly compared to other
dynamical timescales. We shall construct an adiabatic approximation in the form
of a derivative expansion.

Some conditions will be imposed on the radiation. The thermalisation timescale
which governs the return to equilibrium of a small non-thermal fluctuation in
the radiation will be denoted by $\tau_r$ , the evolution timescale of the
inflaton field will denoted by $\tau_\phi$ and the expansion timescale of the
universe by $\tau_a$.  We take
\begin{itemize}
\item $\tau_r\ll\tau_\phi$
\item $\tau_r\ll\tau_a$
\item $T\gg H$
\end{itemize}
The first two conditions are necessary if the system is to remain close to
thermal equilibrium. The second and third conditions allow us to neglect the
expansion of the universe when applying the thermal field theory. Techniques
exist for taking into account both expansion and thermal effects (see
\cite{Boyanovsky:1993xf,Berera:2004kc} 
for example), but typically we expect either the thermal or expansion effects
to predominate, and in this paper we consider the former only.

Consider the
effective field equation,
\begin{equation}
F(x)=-\left.{\delta\Gamma\over\delta\phi_1(x)}\right|_{\phi_a=\phi}
\end{equation}
where $\phi$ is spatially homogeneous and varies slowly about its value
$\phi_t$ at time $t$. We can set $\delta\phi_a=\phi_a-\phi_t$ and expand $F$
by
\begin{equation}
F(x)=\sum_{n=0}^\infty F_n(x)
\end{equation}
where
\begin{equation} 
F_n(x)=-{1\over n!}\int d^4x_1\dots d^4x_n
\left.{\delta^{n+1}\Gamma\over\delta\phi_1(x)
\delta\phi_{a_1}(x_1)\dots\delta\phi_{a_n}(x_n)}\right|_{\phi_a=\phi_t}
\delta\phi_{a_1}(x_1)\dots\delta\phi_{a_n}(x_n)
\end{equation}
The first term $F_0$ represents the part of the field equations which contains 
no derivative terms, and can be expressed as the derivative of an effective
potential which we denote by $V_\phi$. 

The next term $F_1$ contains the derivative terms from the classical action and
the equilibrium self-energy $\Pi_{ab}^\eta({\bf k},t-t_1)$ of the inflaton
perturbations $\eta$. 
According to the maximum time rule, we only need to integrate over times
$t_1<t$. The spatial integral is equivalent to evaluating the self-energy at
zero spatial momentum,
\begin{equation}
F_1(x)=\ddot\phi+3H\dot\phi-
\int_{-\infty}^t dt_1\,\left.\Pi^\eta_{1a}(0,t-t_1)\delta\phi^a(t_1)
\right|_{\phi_a=\phi}\label{nldisp}
\end{equation}
We shall assume that the self-energy introduces a response timescale $\tau$. 
If $\phi$ is slowly varying on the response timescale $\tau$, 
then we can use a simple
Taylor expansion and write
\begin{equation}
\phi(t_1)=\phi(t)+(t_1-t)\dot\phi(t)+\dots
\end{equation}
The inflaton equation of motion including the linear dissipative terms is then
\begin{equation}
\ddot\phi+(3H+\Gamma)\dot\phi+V_\phi=0
\end{equation}
with dissipation coefficient
\begin{equation}
\Gamma=2\int_0^\infty dt'
\,{\rm Re}(\Pi^\eta_{21}(0,t'))\,t'\label{cf},
\end{equation}
where we have used $\Pi_{12}=-\Pi_{21}^*$.

This result is typical of linear response theory \cite{Kapusta:1989tk}. The
range of applicability depends on the assumption that the system remains close
to equilibrium. We also require $\tau<\tau_\phi$ for the existence of a local
dissipation term. If this inequality breaks down, then we can replace the
dissipation coefficient by the non-local expression (\ref{nldisp}), although
the non-local expression can break down when the leading term $F_1$ is not a
uniform approximation over large timescales \cite{Boyanovsky:2003ui}.

\subsection{Simple example}

We shall illustrate the adiabatic approximation with a simple example which
includes a massive boson field
$\chi$ and a massless fermion $\psi$ with an interaction Lagrangian
\begin{equation}
{\cal L}_I(\phi,\chi,\psi)=-\frac12g^2\phi^2\chi^2-
\frac1{\sqrt{2}}h\chi\bar\psi\psi
\end{equation}
The inflaton field $\phi$ is taken to be a slowly varying function of
time.

\begin{center}
\begin{figure}[ht]
\scalebox{0.5}{\includegraphics{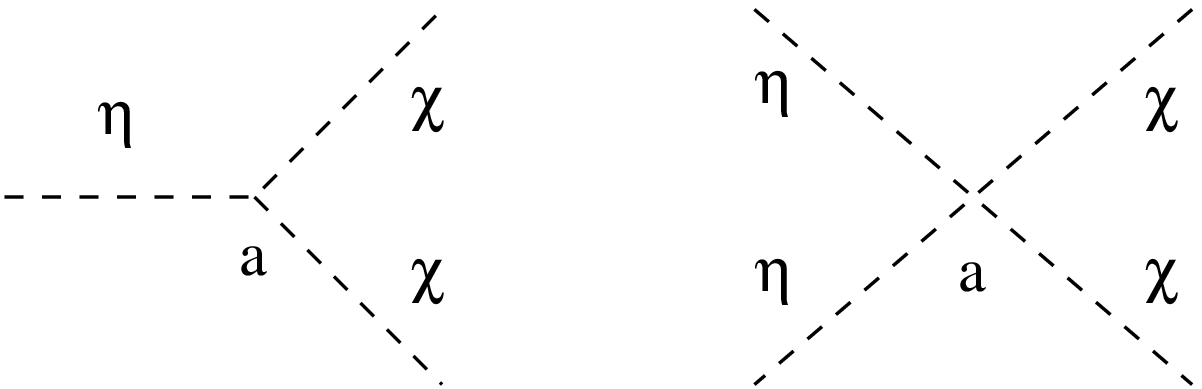}}\par
\hskip .5 true cm  $-2ig^2\phi_a$ \hskip 3
true cm $-2ig^2$
\caption{Vertices for interactions between the fields $\phi$ and $\chi$.}
\label{verts}
\end{figure}
\end{center}

The Feynman diagram vertices are generated in the shifted Schwinger-Keldeysh
formalism by
\begin{equation}
{\cal L}'_I(\phi_a,\eta_a,\chi_a,\psi_a)=
{\cal L}_I(\phi_1+\eta_1,\chi_1,\psi_1)
-{\cal L}_I(\phi_2+\eta_2,\chi_2,\psi_2)
\end{equation}
The vertices for interactions between $\eta$ and $\chi$ are shown in figure
\ref{verts}.  We are interested in the situation where $\chi$ as a heavy field,
with mass $m_\chi\sim g\phi$. We shall therefore regard the first vertex as
$O(g)$.

\begin{center}
\begin{figure}[ht]
\scalebox{0.5}{\includegraphics{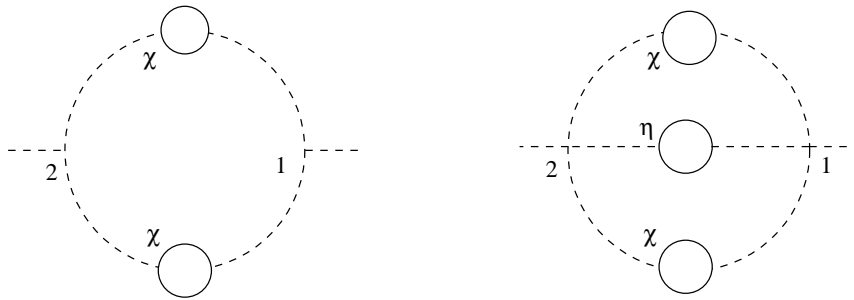}}
\caption{Contributions to the $\eta$ self-energy of order $g^2$ (left) and
$g^4$ (right).}
\label{simple}
\end{figure}
\end{center}

The dissipation coefficient can be obtained from eq. (\ref{cf}).  The
contribution
to the self-energy of the $\eta$ field at order $g^2$ is given by  the first
diagram in figure \ref{simple} with two $\chi$ propagators,
\begin{equation}
\Gamma=4g^4\phi^2\,{\rm Im}\int {d^3k\over (2\pi)^3}\int_0^\infty dt'\,
\left( G^\chi_{21}(t')\right)^2 t'.\label{fc}
\end{equation}
This can also be written in terms of the spectral function $\rho_\chi$, using
(\ref{bspec}),
\begin{equation}
G^\chi_{21}(t')=\int_{-\infty}^\infty {d\omega\over 2\pi}i(1+n)\rho_\chi
\,e^{-i\omega t'}.
\end{equation}
After applying the identity
\begin{equation}
{\rm Im}\int_0^\infty dt'\,t'e^{-i\omega t'}=
-\pi\,\delta'(\omega),
\end{equation}
we get
\begin{equation}
\Gamma=-2g^4\phi^2\,
\int {d^3k\over (2\pi)^3}\int_0^\infty {d\omega\over 2\pi}\,
\rho_\chi^2\,n'.\label{simplegam}
\end{equation}

There are two contrasting limits in which the result can be replaced by an
approximation. First there is the weakly coupled limit with small $h$ and
fixed $T$. The energy integral is dominated by the point $\omega=\omega_k$,
\begin{equation}
\omega_k=(k^2+m_\chi^2)^{1/2},
\end{equation}
which lies close to two poles in the spectral function (\ref{specb}). These two
poles are at $\omega=\omega_k\pm i\tau_\chi^{-1}$ (possibly on a different
Riemann sheet due to branch cuts), where $\tau_\chi$ is the relaxation time
for the $\chi$ boson. The integrand can be expanded about
$\omega=\omega_k$ to obtain the
result first derived by Hosoya and Sakagami \cite{hosoya84},
\begin{equation}
\Gamma\approx g^4\phi^2\,\beta\int {d^3k\over (2\pi)^3}
 {\tau_\chi \over2\omega_k^2}n(n+1).
\end{equation}
The contribution to $\tau_\chi$ due to a single fermion loop is of order
$h^{-2}$, as we shall see in a later section, so that $\Gamma=O(g^2h^{-2})$.

The low temperature limit $\beta m_\chi\to\infty$ for fixed coupling $h$ gives
a quite different approximation. In this case, exponential terms control the
energy and momentum integrals and the dominant contributions come from
$\omega\ll m_\chi$ and $k\ll m_\chi$. According to equation (\ref{specb}), the
spectral function can be replaced by
\begin{equation}
\rho_\chi\approx {4\Gamma_\chi\over m_\chi^3}
\end{equation}
where $\Gamma_\chi$ is the low-energy decay width of the $\chi$ boson.
The dissipation coefficient given by equation (\ref{fc}) becomes
\begin{equation}
\Gamma\sim{g^4T^3\phi^2\over m_\chi^6}
\int {d\omega\over 2\pi}\, {d^3k\over(2\pi)^3}\,
32\beta^4\Gamma_\chi^2\,n(n+1)
\end{equation}
The magnitude of this term depends on the behaviour of the vacuum polarisation
at small $\omega$. We shall give results for $\Gamma_\chi$ in a later section.
The
fermion contribution to the vacuum polarisation gives a contribution
$\Gamma_\chi\sim h^2K^2m_\chi^{-1}$, and $\Gamma\propto T^7$. The dominant
contribution to
$\Gamma$ at low temperature comes from light scalar interactions, which we
shall consider in section III.

\subsection{Zero-temperature dissipation terms}

The simple example of the previous section suggests that the dissipation
coefficient in the inflaton equation vanishes at order $g^2$ in the zero
temperature limit for a non-expanding universe. This contradicts some previous
results \cite{Morikawa:1984dz,Berera:2001gs,berera02,Berera:2004kc}, but it is
consistent with the conclusions of Boyanovsy et al. \cite{Boyanovsky:1994me},
who found that their zero-temperature dissipation term could not be localised.
We shall therefore examine this issue further.

Consider the simple model again, with the $O(g^2)$ contribution to the
dissipation coefficient
\begin{equation}
\Gamma=4g^4\phi^2\,{\rm Im}\int{d^3k\over (2\pi)^3}\int_0^\infty dt'
G_{21}^\chi(t')^2\,t'\label{disp}
\end{equation}
In the vacuum, $G_{21}^\chi$ is the Wightman function, and includes corrections
due to interactions with light fields. The spectral representation of the
Wightman function is non-vanishing only when the energy lies in the physical
particle spectrum,
\begin{equation}
G_{21}(\omega)=i\theta(\omega)\rho_\chi(\omega).
\end{equation}
The dissipation coefficient becomes
\begin{equation}
\Gamma=4g^4\phi^2\,{\rm Im}\int{d^3k\over (2\pi)^3}
\int_k^\infty{d\omega_2\over 2\pi}\int_k^\infty{d\omega_2\over 2\pi}
\int_0^\infty dt'\,\rho_\chi(\omega_1)\rho_\chi(\omega_2)
\,t'e^{-i(\omega_1+\omega_2-i\epsilon)}
\end{equation}
which always vanishes.

In striking contrast, non-zero values of the dissipation coefficient can be
obtained by using an approximation for the $\chi$ propagator, which we shall
refer to as the exponential decay approximation,
\begin{equation}
G_{MS}(t)\approx{i\over 2(\omega_k-i\Gamma_\chi)}
e^{-i(\omega_k-i\Gamma_\chi) t}\qquad  t>0,\label{wvf}
\end{equation}
where $\omega_k=(k^2+m_\chi^2)^{1/2}$ and $\Gamma_\chi$ is the $\chi$ decay
width. This approximation gives a non-zero value for the dissipation
coefficient $\Gamma_{MS}$ from eq. (\ref{disp}),
\begin{equation}
\Gamma_{MS}=4g^2\phi^2{\rm Im}\int{d^3k\over (2\pi)^3}
{1\over (\omega_k-i\Gamma_\chi)^4}\label{gammams}
\end{equation}
This result was first obtained by Morikawa and Sasaki \cite{Morikawa:1984dz}.

The difference between to two results is due to the energy dependence of the
width $\Gamma_\chi$ and the consequent failure of the
exponential decay approximation for the zero-temperature propagator. The
simplest example is when the $\chi$ propagator has a self-energy
contribution $\Pi$ from a massless boson loop, which gives
\begin{equation}
\Pi={\gamma^2\over2\pi}\ln(k^2-\omega^2-i\epsilon)
\end{equation}
where $\gamma$ is a constant. The Feynman green function
\begin{equation}
G_F(\omega)=(-\omega^2+\omega_k^2+\Pi)^{-1}
\end{equation}
has branch cuts above and below the real $\omega$ axis which push the particle
pole off the first Riemann sheet. After some contour
manipulation,
\begin{equation}
G_F(t)=\int_k^\infty {d\omega\over 2\pi}
\left\{(-\omega^2+\omega_k^2-2i\Gamma_\chi\omega_k)^{-1}
-(-\omega^2+\omega_k^2+2i\Gamma_\chi\omega_k)^{-1}\right\}
e^{-i\omega t}\qquad  t>0
\end{equation}
where $\Gamma_\chi=\gamma^2/(2\omega_k)$ and the real part of the vacuum
polarisation has been dropped. The result can also be obtained directly from
the spectral representation of the Wightman function.

The energy integral can be done analytically,
\begin{equation}
G_F(t)={i\over 2(\omega_k-i\Gamma_\chi)}e^{-i(\omega_k-i\Gamma_\chi) t}
+f(\omega_k-i\Gamma_\chi,t)
-f(\omega_k+i\Gamma_\chi,t)\label{wft}
\end{equation}
with an error function part
\begin{equation}
f(\omega,t)={e^{i\omega t}\over 4\pi\omega}E_1(i(k+\omega)t)
-{e^{-i\omega t}\over 4\pi\omega}E_1(i(k-\omega)t)
\end{equation}
We recognise the first term in (\ref{wft}) as the earlier approximation
(\ref{wvf}). This term is a good approximation to the full expression for early
times up to $t\sim\tau'_\chi$, where
\begin{equation}
\tau'_\chi=\Gamma_\chi^{-1}\log{m_\chi^2\over\Gamma_\chi^2},
\end{equation}
when the error functions begin to dominate and the propagator decays as a power
law. When (\ref{wft}) is substituted
into the formula for the dissipation coefficient
(\ref{jeq}), we find
\begin{equation}
\Gamma=4g^4\phi^2{\rm Im}\int {d^3k\over (2\pi)^3}\int_0^\infty
dt'G_F(t')^2t'=0,
\end{equation}
due to a cancellation between the exponential and the error function terms.

A similar argument shows that the equilibrium propagator also has a power law
decay at large time. However, it may be possible that an exponential decay
law is the best representation the non-equilibrium propagators in the
dissipation term, as has been suggested by Lawrie
\cite{lawrie01,Lawrie:2002zd}. 
There is some support for this suggestion from numerical evidence that the
real-time spectral function in two dimensions has an exponential decay which
extends well beyond the relaxation time \cite{Aarts:2001qa}. The implication
would be that $\Gamma_{MS}$ may be applicable if the inflaton evolution
timescale $\tau_\phi$ lies between the relaxation time $\tau_\chi$ and the
thermalisation time of the heavy field. This issue awaits further
investigation.

\subsection{Non-local dissipation terms}

The zero-temperature propagator has a well-defined relaxation time
$\tau_\chi=\Gamma_\chi^{-1}$, but fails to produce a local dissipation term in
the inflaton equation due to the non-exponential terms. We shall now
investigate the form of the non-local
dissipation term. We begin by introducing a transform of the
inflaton field
\begin{equation}
\phi(\omega)=\int_{-\infty}^\infty dt'\phi(t')e^{i\omega(t'-t)}
\end{equation}
This substitution gives the linear dissipation term in the equation of motion,
\begin{equation}
F_1^{non-loc}=-4g^4\phi^2{\rm Im}\int {d^3k\over (2\pi)^3}\int_0^\infty
{d\omega_1\over
2\pi}{d\omega_2\over 2\pi}{d\omega_3\over 2\pi}
\rho_\chi(\omega_1)\rho_\chi(\omega_2)
\,i(\omega_1+\omega_2-\omega_3-i\epsilon)^{-1}
\phi(\omega_3)\label{jeq}
\end{equation}
A more suggestive way of rewriting (\ref{jeq}) is to set
\begin{equation}
F_1^{non-loc}=2g^2{\rm Im}\left(g^2\phi^2 K(\partial_t)\phi\right).
\end{equation}
where the kernel
\begin{equation}
K(z)=-{\rm Im}\int {d^3k\over (2\pi)^3}\int_0^\infty 
{d\omega_1\over 2\pi}
{d\omega_2\over 2\pi}
\rho_\chi(\omega_1)\rho_\chi(\omega_2)\,i(\omega_1+\omega_2+iz+i\epsilon)^{-1}
\label{keq}
\end{equation}
This would be a local derivative expansion if $K(z)$ was an analytic function.
Although $K(z)$ is finite at $z=0$, there is a branch cut and $K(z)$ is
non-analytic in the neighbourhood of the origin. 

We can isolate a leading order logarithmic term in $K(z)$ by taking {\it six}
derivatives with respect to $z$ and then taking a small $z$ approximation. 
For small $z$,
\begin{equation}
\partial_z{}^6K(z)\sim -120\rho_\chi(0)^2\int {d^3k\over (2\pi)^3}\int_k^\infty
{d\omega_1\over 2\pi}
\int_k^\infty {d\omega_2\over 2\pi}
\,(\omega_1+\omega_2+iz)^{-6}\sim 
{-i\over 32\pi^4}\rho_\chi(0)^2z^{-1}
\end{equation}
where $\rho_\chi(0)\approx 4\Gamma_\chi/m_\chi^3$. (At zero temperature,
$\rho_\chi$
vanishes for $\omega<k$, and it is important to take the
limit $\omega\to k$ and then take the limit $k\to 0$). Hence,
\begin{equation}
2{\rm Im}K(z)\sim K_0+K_2z^2+K_4z^4-
{1\over 6\pi^2}{\Gamma_\chi^2\over m_\chi^6}z^4\ln z
\end{equation}
where $K_0\dots K_4$ are constants. The important term for dissipation is the
logarithm, which appears the field equations as
\begin{equation}
\ddot\phi+3H\dot\phi+V_{\phi}
-{g^2\over 3\pi^2}{\Gamma_\chi^2\over m_\chi^6}g^2\phi^2\,
\partial_t^4\ln (\partial_t)\,\phi=0\label{fone}
\end{equation}
This dissipative term represents the leading order adiabatic behaviour of the
full expression (\ref{jeq}). 

The non-local operator $\ln(\partial_t)$ is fixed by linearity and causality as
discussed in \cite{Moss:2004dp}. It can be expressed in integral form,
\begin{equation}
\ln\left(\partial_t\right)f
=-\int_{-\infty}^t\ln\left(e^\gamma(t-t')\right)\partial_{t'}f(t')dt'
\end{equation}
where $\gamma$ is Euler's constant. The integral can be evaluated
for a broad class of functions, for example
\begin{equation}
\ln\left(m^{-1}\partial_t\right)\theta(t)\sin(m t)=
{\rm Si}(m t)\cos(m t)
-{\rm Ci}(m t)\sin(m t)
\end{equation}
where ${\rm Si}$ and ${\rm Ci}$ are sine and cosine integrals. (This can be
used to apply eq. (\ref{fone}) to the decay of an oscillating field
$\phi=A\sin(mt)$, where the amplitude is a slowly varying function of time 
\cite{Chang:2004xs}.)

\section{Self-energy}

We shall examine the dissipative terms in the equations of motion for the
inflaton with a range of interactions which are designed to be embedded easily
in supersymmetric models. In this section we shall calculate the self energy
for bosonic and fermionic massive $\chi$ fields interacting with light bosonic
and fermionic fields of mass $m_y$. 

The specific interaction terms for the $O(h^2)$ contributions to the
self-energy are
\begin{enumerate}
\item For $\chi\to2 y$
\begin{equation}
{\cal L}_Y=-\frac1{\sqrt{2}}
gh\phi(y^2\chi^\dagger+y^{\dagger2}\chi)\label{cyy}
\end{equation}
\item For $\chi\to2\tilde y$
\begin{equation}
{\cal L}_Y=-
\frac1{\sqrt{2}}h(\chi\bar\psi_yP_L\psi_y+\chi^\dagger\bar\psi_yP_R\psi_y)
\end{equation}
\item For $\tilde\chi\to y\tilde y$
\begin{equation}
{\cal L}_Y=-
{\sqrt{2}}h(y\bar\psi_\chi P_L\psi_y+y^\dagger\bar\psi_\chi P_R\psi_y)
\end{equation}
\end{enumerate}
where $P_L=\frac12(1+\gamma_5)$. The fermions used in the supersymmetric models
are all Majorana. 

\subsection{Interaction 1: $\chi\to2y$}

\begin{center}
\begin{figure}[ht]
\scalebox{0.5}{\includegraphics{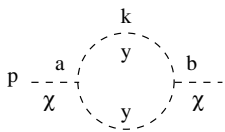}}
\caption{Order $h^2$ contribution to the boson self-energy $\Pi_{ab}$.}
\label{figvac3}
\end{figure}
\end{center}

The leading order contribution to the self-energy from the light boson
loop is given by figure \ref{figvac3}
We shall calculate the dissipative part of the self-energy, and denote by
$\alpha_y$ the function defined in equation (\ref{pa}),
\begin{equation}
\alpha_y=i(e^{\beta\omega}-1)\Pi_{12}
\end{equation}
The loop diagram gives,
\begin{equation}
\alpha_y=-g^2h^2\phi^2(e^{\beta\omega}-1)\int {d^4K\over(2\pi)^4}
G_{12}(K)G_{12}(P-K)
\end{equation}
A spectral representation (\ref{fspec}) can now be applied,
\begin{equation}
\alpha_y=\frac12
h^2(e^{\beta\omega}-1)\int {dk^0\over 2\pi}{dk^{0\prime}\over 2\pi}
{d^3k\over(2\pi)^3} n(k^0)n(k^{0\prime})
\delta(\omega-k^0-k^{0\prime})\,
\rho({\bf k},k^0)\rho({\bf k-p},k^{0\prime})
\end{equation}
We can use the free bosonic spectral function  given in the appendix
(\ref{freespecb}). The integral reduces to a different form for different
ranges of energy. Let
\begin{equation}
\omega_c=(4m_y^2+p^2 )^{1/2}.
\end{equation}
In the $\omega>\omega_c$ case, the integral reduces down to
\begin{equation}
\alpha_y=2\pi\,g^2h^2\phi^2(e^{\beta\omega}-1)\int
{d^3k\over(2\pi)^3}
{1+2n(\omega_k)\over 4\omega_k\omega_{p-k}}
\,\delta(\omega-\omega_k-\omega_{p-k}).
\end{equation}
This integral can be done using a substitution
\begin{eqnarray}
y&=&\omega_k+\omega_{p-k}\\
x&=&\omega_k-\omega_{p-k}.
\end{eqnarray}
The range of integration becomes $y>\omega_c$ and $-x(P)<x<x(P)$, where
\begin{equation}
x(P)=p\left(1+{4m_y^2\over P^2}\right)^{1/2}.\label{xp}
\end{equation}
In the end, we get
\begin{equation}
\alpha_y={h^2\over 8\pi}g^2\phi^2\left(1+{4m_y^2\over P^2}\right)^{1/2}
+{h^2\over 4\pi}g^2\phi^2{T\over p}
\ln\left({1-e^{-\beta(\omega+x(P))/2}\over 
1-e^{-\beta(\omega-x(P))/2}}\right)\qquad \omega>\omega_c.
\end{equation}
Repeating the analysis for $0<\omega<\omega_c$ gives $\alpha=0$ for
$p<\omega<\omega_c$ and
\begin{equation}
\alpha_y={h^2\over 4\pi}g^2\phi^2{T\over p}
\ln\left({1-e^{-\beta(x(P)+\omega)/2}\over 
1-e^{-\beta(x(P)-\omega)/2}}\right)\qquad 0<\omega<p.
\end{equation}
These are quite a complicated expressions, but they can be interpreted simply
in certain limits. At zero temperature only the first term from the
$\omega>\omega_c$ result survives. Since $\alpha$ becomes the imaginary part of
the vacuum polarisation, this term is related to the particle production
probability and $\omega_c$ is the energy threshold for particle production. 

At the other extreme, for high temperatures $\alpha=O(h^2m_\chi^2)$. This
result can be checked by taking the Hard Thermal Loop approximation
\cite{lebellac}, which when applied here predicts that the imaginary part of
the scalar self-energy should have no $O(T^2)$ contribution. This contrasts
with the vacuum polarisation obtained from other types of vertex. Gleiser and
Ramos \cite{gleiser94} considered the scalar vaccum polarisation due to
quadratic interactions, and their results suggest that  $h^2\chi^2y^2$
vertices would give $\alpha=O(h^4T^2)$ in the high temperature limit. We shall
restrict ourselves to temperatures $T<h^{-1}m_\chi$ for which the $O(h^4T^2)$
terms can be neglected.

For light fields we can approximate the result by taking the massless limit
$m_y=0$,
\begin{equation}
\alpha_y={h^2\over 8\pi}g^2\phi^2\,\theta(\omega-p)
+{h^2\over 4\pi}g^2\phi^2
{T\over p}\ln\left({1-e^{-\beta(\omega+p)/2}\over 
1-e^{-\beta|\omega-p|/2}}\right).\label{alb}
\end{equation}
This approximation breaks down at the point $\omega=p$, but remains integrable.

\subsection{Interaction 2: $\chi\to2\tilde y$}

\begin{center}
\begin{figure}[ht]
\scalebox{0.5}{\includegraphics{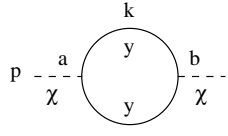}}
\caption{Order $h^2$ fermionic contribution to the boson self-energy.}
\label{figvac1}
\end{figure}
\end{center}

The leading order contribution to the self-energy from a fermion loop
is given by figure \ref{figvac1}.
We shall calculate the dissipative part and denote this contribution by
$\alpha_{\tilde y}$.
The fermion loop diagram gives,
\begin{equation}
\alpha_{\tilde y}=h^2(1-e^{\beta\omega})\int {d^4K\over(2\pi)^4}
{\rm tr}(S_{12}(K)S_{21}(K-P)).
\end{equation}
A spectral representation (\ref{fspec}) can now be applied. In the
$\omega>\omega_c$ case, the integral reduces down to
\begin{equation}
\alpha_{\tilde y}=-2\pi h^2\int {d^3k\over(2\pi)^3}
{1-2n(\omega_k)\over 4\omega_k\omega_{p-k}}(P^2+4m_y^2)
\,\delta(\omega-\omega_k-\omega_{p-k}).
\end{equation}
This integral can be done using the same substitution as before, resulting in
\begin{equation}
\alpha_{\tilde y}=-{h^2\over 8\pi}P^2\left(1+{4m_y^2\over P^2}\right)^{3/2}
-{h^2\over 4\pi}{T\over p}P^2\left(1+{4m_y^2\over P^2}\right)
\ln\left({e^{-\beta(\omega+x(P))/2}+1\over 
e^{-\beta(\omega-x(P))/2}+1}\right)\qquad \omega>\omega_c.
\end{equation}
Repeating the analysis for $0<\omega<\omega_c$ gives
\begin{equation}
\alpha_{\tilde y}=-{h^2\over 4\pi}{T\over p}P^2\left(1+{4m_y^2\over P^2}\right)
\ln\left({e^{-\beta(x(P)+\omega)/2}+1\over 
e^{-\beta(x(P)-\omega)/2}+1}\right)\qquad 0<\omega<p.
\end{equation}
The fermion loop, like the boson loop, gives $\alpha=O(h^2m_\chi^2)$ in the
high temperature limit.

In the massless $m_y=0$ limit,
\begin{equation}
\alpha_{\tilde y}=-{h^2T^2\over 8\pi}(\beta P)^2\,\theta(\omega-p)
-{h^2T^2\over 4\pi}{1\over \beta p}(\beta P)^2
\ln\left({e^{-\beta(\omega+p)/2}+1\over 
e^{-\beta|\omega-p|/2}+1}\right).\label{alf}
\end{equation}
The momenta have been scaled to display the dependence on the temperature in
the low temperature limit. Note that, at low temperatures, the vacuum
polarisation of the $\chi$ field is dominated by the bosonic loop
contribution.

\subsection{Interaction 3: $\tilde\chi\to y\tilde y$}

\begin{center}
\begin{figure}[ht]
\scalebox{0.5}{\includegraphics{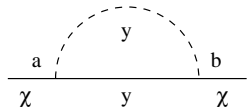}}
\caption{Order $h^2$ contribution to the fermion self-energy $\Sigma_{ab}$.}
\label{figvac2}
\end{figure}
\end{center}

The leading order contribution to the fermion self-energy is given by figure 
\ref{figvac2}. 
We shall calculate the function $\tilde\alpha$, given by equation (\ref{sa}),
\begin{equation}
\tilde\alpha=-i(e^{\beta\omega}+1)\Sigma_{12}.
\end{equation}
From the diagram we have,
\begin{equation}
\tilde\alpha=2h^2(e^{\beta\omega}+1)
\int {d^4K\over (2\pi)^4}G_{12}(K)S_{12}(P-K).
\end{equation}
We proceed as before. For $\omega>\omega_c$,
\begin{equation}
\tilde\alpha=4\pi h^2\int {d^3k\over (2\pi)^3}
\left(1-n(\omega_k)+\tilde n(\omega_{p-k})\right)
(-\gamma^\mu (P_\mu-K_\mu)+m_y)\delta(\omega-\omega_k-\omega_{p-k}).
\end{equation}
The substitution of the new variables $x$ and $y$ can be used again. The
gamma-matrix terms become
\begin{equation}
-\gamma^\mu (P_\mu-K_\mu)=
\frac12\sigma_0-{x\over 2p}\sigma_1,
\end{equation}
where $\sigma_0$ and $\sigma_1$ are defined in (\ref{sig}).
The integral gives
\begin{equation}
\tilde\alpha=\tilde\alpha_0\sigma_0+\tilde\alpha_1\sigma_1,
+\tilde\alpha_m m_\chi,
\end{equation}
with
\begin{equation}
\tilde\alpha_0={m_{\tilde\chi}\over 2m_y}\tilde\alpha_m=
{h^2\over 8\pi}
\left(1+{4m_y^2\over P^2}\right)^{1/2}
+{h^2\over 8\pi}{T\over p}\ln\left({1-e^{-\beta(\omega+x(P))}\over 
1-e^{-\beta(\omega-x(P))}}\right)
\qquad\omega>\omega_c.\label{alzero}
\end{equation}
and
\begin{eqnarray}
\tilde\alpha_1&=&{h^2\over 8\pi}\left(1+{4m_y^2\over P^2}\right)^{1/2}
{T\over p}\ln\left({1-e^{-\beta(\omega+x(P))}\over 
1-e^{-\beta(\omega-x(P))}}\right)\nonumber\\
&&+{h^2\over 8\pi}\left({T\over p}\right)^2
\left({\rm Li_2}(e^{-\beta(\omega-x(P))/2})-
{\rm Li_2}(-e^{-\beta(\omega-x(P))/2})-
{\rm Li_2}(e^{-\beta(\omega+x(P))/2})+
{\rm Li_2}(e^{-\beta(\omega+x(P))/2})\right),\label{alone}
\end{eqnarray}
where $x(p)$ was given in eq. (\ref{xp}) and
\begin{equation}
{\rm Li_2}(z)=\sum_{n=1}^\infty{z^n\over n^2}
\end{equation}
is the dilogarithm function.

It is worth noting at this stage that the relaxation time of the $\tilde\chi$
field does not depend on the $\sigma_1$ terms. This is helpful because
dissipation coefficient in the small coupling limit is largely determined by
the
relaxation time. If we use eq. (\ref{fgam}) 
for the thermal width $\Gamma_{\tilde\chi}$,
with $m= m_{\tilde \chi}=g\phi$, we get
\begin{equation}
2\omega_p\Gamma_{\tilde\chi}={h^2\over 4\pi}
g^2\phi^2\left(1+{2m_y\over m_{\tilde\chi}}\right)^{3/2}
\left(1-{2m_y\over m_{\tilde\chi}}\right)^{1/2}
+{h^2\over 4\pi}g^2\phi^2{T\over p}
\left(1+{2m_y\over m_{\tilde\chi}}\right)
\ln\left({1-e^{-\beta(\omega+x(P))}\over 
1-e^{-\beta(\omega-x(P))}}\right).
\end{equation}
We are interested mostly in very light fields $y$, for which we can use the
limit $m_y=0$ when
\begin{equation}
2\omega_p\Gamma_\chi={h^2\over 4\pi}
g^2\phi^2\theta(\omega-p)
+{h^2\over 4\pi}g^2\phi^2{T\over p}
\ln\left({1-e^{-\beta(\omega+p)}\over 
1-e^{-\beta|\omega-p|}}\right).\label{gamp}
\end{equation}
This is identical to the thermal width of the massive boson fields when both
the light boson and light fermion loops are included.

\section{More dissipation terms}

The calculation of the frictional terms in the inflaton equation of motion now
needs to be extended to a more general class of models. The first 
generalisation we do is to take into account the coupling of the light fields
to the inflaton which arises in eq. (\ref{cyy}).  The other generalisation is
to consider dissipation terms which arise due to interactions between the
inflaton and massive fermionic fields.

The interactions which we consider are all contained in a supersymmetric theory
which has three superfields $\Phi$, $X$ and $Y$ with superpotential,
\begin{equation}
W=\frac{g}{\sqrt{2}}\Phi X^2-\frac{h}{\sqrt{2}}XY^2.
\end{equation}
The scalar components of the superfields are $\phi$, $\chi$ and $y$
respectively. This superpotential can easily be modified to produce a hybrid
inflationary model by adding additional terms or by adding a $D$-term to the
potential, neither of which affects the dissipation terms.

The interaction terms responsible for the inflaton decay are 
\begin{enumerate}
\item For $\phi\to\chi$
\begin{equation}
{\cal L}_I=-g^2\phi^2\chi^\dagger\chi;
\end{equation}
\item For $\phi\to\tilde\chi$
\begin{equation}
{\cal L}_I=-\frac12g\phi\,\bar\psi_\chi\psi_\chi.
\end{equation}
\end{enumerate}
The scalar  interaction vertices are given by figure \ref{verts}.

\subsection{Dissipation with an intermediate boson: $\phi\to2y$ and 
$\phi\to 2\tilde y$}

\begin{center}
\begin{figure}[ht]
\scalebox{0.5}{\includegraphics{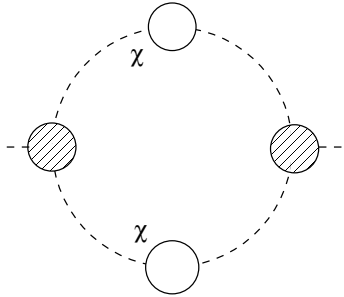}}
\caption{Contribution the $\eta$ self-energy.}
\label{vacuum}
\end{figure}
\end{center}

\begin{center}
\begin{figure}[ht]
\scalebox{0.5}{\includegraphics{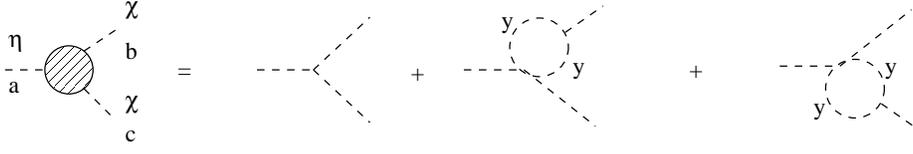}}
\caption{Contributions to the $\eta$ vertex up to order $g^2h^2$ .}
\label{vertex}
\end{figure}
\end{center}

Consider the dissipation coefficient given by eq. (\ref{cf}) with the
equilibrium
self energy diagram shown in figure
\ref{vacuum}. The dissipation coefficient in this case will be denoted by
$\Gamma(\phi\to 2y/2\tilde y)$. It is given to order $g^2$ by
\begin{equation}
\Gamma(\phi\to 2y/2\tilde y)={\rm Im}\,
\int {d^3 k\over (2\pi)^3}{d\omega_1\over 2\pi}{d\omega_2\over 2\pi}
\Gamma_2{}^{ab}\Gamma_1{}^{a'b'}
G^\chi_{aa'}(\omega_1)G^\chi_{bb'} (\omega_2)2\pi i\
\partial_{\omega_1}\delta(\omega_1+\omega_2),\label{cgam}
\end{equation}
where the $\Gamma_{abc}$ are vertex factors. The vertex factors now include
contributions from the light
fields, as shown in figure (\ref{vertex}). The vacuum loop inside the vertex
factor is similar to
the leading order correction to the $\chi$ self-energy, which allows
to write the vertex factor in terms of $\alpha_y$, as
\begin{equation}
\Gamma_2{}^{ab}(\omega,\omega_1,\omega_2)=
-2ig^2\phi \,c_{2}{}^{ab}
+\phi^{-1}(1+n)\alpha_y(\omega_1)c_{2}{}^{a\bar b}
+\phi^{-1}(1+n)\alpha_y(\omega_2)c_{2}{}^{\bar a b}
+O(h^4)\label{vert}
\end{equation}
where $c_{a}{}^{bc}=\delta_{a}{}^{b}\delta_{a}{}^{c}(2\pi)^4\delta(K+K_1+K_2)$, 
$\bar 1=2$ and $\bar 2=1$.

The vertex factor can be substituted into eq. (\ref{cgam}). We define
\begin{equation}
\sigma=2\,{\rm Re}\,G_{11}=-2\,{\rm Re}\,G_{22}
\end{equation}
and use the spectral representation $G_{21}=i(1+n)\rho_\chi$. The integral
reduces
down to
\begin{equation}
\Gamma(\phi\to 2y/2\tilde y)=\int{d^3 k\over
(2\pi)^3}\int_0^\infty{d\omega\over 2\pi}
\left( 4g^4\phi^2\,\rho_\chi^2+
4g^2\,\rho_\chi\,\alpha_y\,\sigma+
\frac12\phi^{-2}\alpha_y^2\sigma^2\right)n'.\label{gin}
\end{equation}
The first term recovers our result (\ref{simplegam}) from section IIIB, with an
extra factor of two because $\chi$ is now a complex field.

\begin{center}
\begin{figure}[ht]
\scalebox{1.0}{\includegraphics{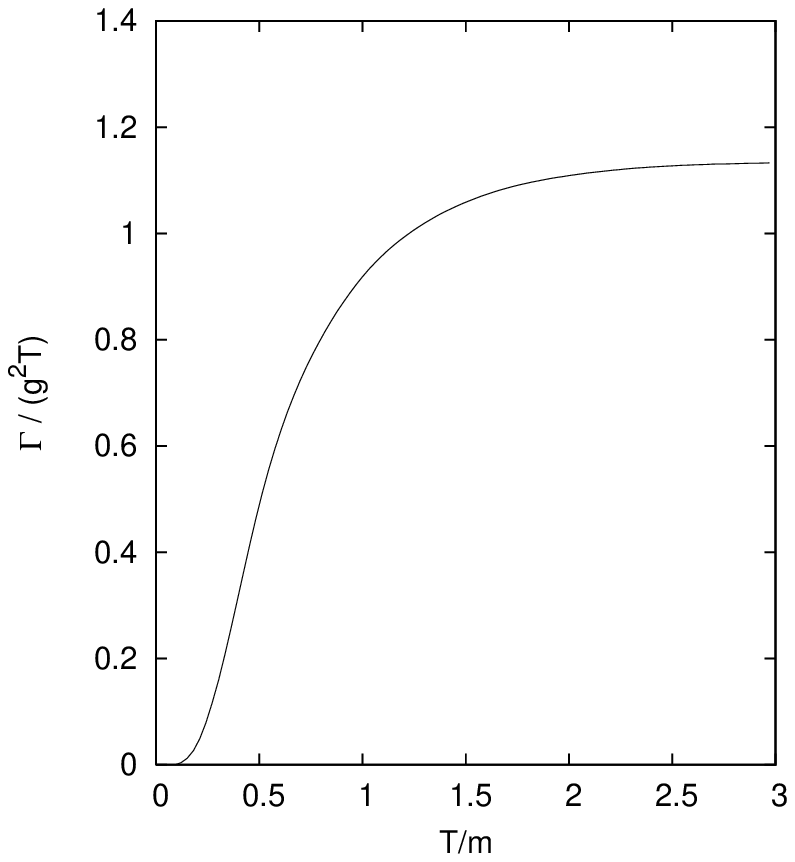}\includegraphics{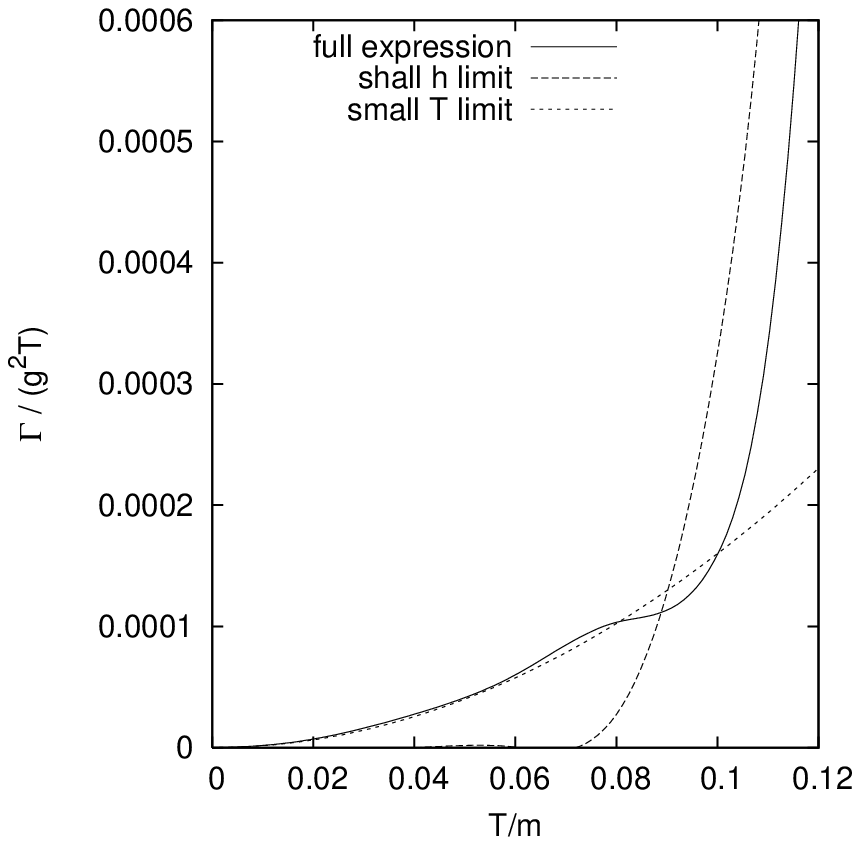}}
\caption{The dissipation coefficient $\Gamma(\phi\to 2y/2\tilde y)$ plotted as
a
function
of temperature for $h=0.79$. The plot on the right shows the low temperature
region where $\Gamma(\phi\to 2y/2\tilde y)\propto T^3$ and the small $h$
approximation
breaks down. }
\label{figgplot}
\end{figure}
\end{center}
Numerical values for the dissipation coefficient are plotted in figure
\ref{figgplot}. The small $h$ approximation works reasonably well even for
$h\sim 1$, except in the low temperature regime.

For small coupling, $h$, the integrand is dominated by poles in
the spectral function. The last two terms in eq. (\ref{gin}) are higher order
in $h$ and can be dropped.  The first term gives
\begin{equation}
\Gamma(\phi\to 2y/2\tilde y)\approx g^4\phi^2\,\beta\int {d^3k\over (2\pi)^3}
 {\tau_\chi \over\omega_k^2}n(n+1).
\end{equation}
where $\tau_\chi$ is the relaxation time of the $\chi$ field. For 
$h^{-1}m_\chi>T$, we can use eqs. (\ref{alb}) and (\ref{alf}) for the boson
and fermion loop contributions to the self-energy,
\begin{equation}
\tau_\chi={2\omega_k\over 
\alpha_y(\omega_k,k)+\alpha_{\tilde y}(\omega_k,k)}
\end{equation}
When this is combined with a high temperature approximation, the dissipation
coefficient becomes
\begin{equation}
\Gamma(\phi\to 2y/2\tilde y)\approx 0.691\,
{g^2\over h^2}\,T
\qquad h^{-1}m_\chi> T\gg m_\chi.
\end{equation}
This is shown in figure \ref{figgplot}.

In the low temperature limit the integral (\ref{gin}) is dominated by small $k$
and $\omega$. We can approximate $\rho_\chi\approx2\alpha/m_\chi^4$ and
$\sigma\approx 2/m_\chi^2$. The leading order behaviour is determined by the
self-energy $\alpha_y$,
\begin{equation}
\Gamma(\phi\to 2y/2\tilde y)\approx{g^2T^3\phi^2\over m_\chi^8}
\int {d\omega\over 2\pi}\, {d^3k\over(2\pi)^3}\,
34\beta^4\alpha_y^2\,n'.
\end{equation}
We can use the value of $\alpha_y$ given in eq. (\ref{alb}), to get
\begin{equation}
\Gamma(\phi\to 2y/2\tilde y)\approx
4.0\times 10^{-2}\,g^2h^4\left({g\phi\over m_\chi}\right)^4{T^3\over m_\chi^2}
\qquad T\ll m_\chi.
\end{equation}
The value of the coefficient has been found by numerical integration.

\subsection{Dissipation from an intermediate fermion: $\phi\to y\tilde y$}

The contribution to th dissipation
coefficient from the decay $\phi\to\tilde\chi\to y\tilde y$ is given by
\begin{equation}
\Gamma(\phi\to y\tilde y)=
g^2\,{\rm Im}\int{d^3 k\over (2\pi)^3}\int_0^\infty dt'
\,{\rm tr}(S_{21}^\chi(t')\widetilde S_{21}^\chi(t'))t'
\end{equation}
where the fermion two-point functions and conjugates where defined following
eq. (\ref{fprop}). Using the spectral representation of the two-point functions
in eq. (\ref{fspec}) gives,
\begin{equation}
\Gamma(\phi\to y\tilde y)=\frac12g^2\int{d^3 k\over (2\pi)^3}\int_0^\infty
{d\omega\over 2\pi}
\,{\rm tr}(\tilde\rho_\chi^2)\,\tilde n'.
\end{equation}
In the small coupling limit we can expand the integrand
about the poles of the spectral function using (\ref{fgam}),
\begin{equation}
\Gamma(\phi\to y\tilde y)\approx g^2m^2_{\tilde\chi}\beta\,\int{d^3 k\over
(2\pi)^3}\,{\tau_\chi\over\omega_k^2}\,
\tilde  n(\omega_k)(1-\tilde n(\omega_k))\,.
\end{equation}
The relaxation time $\tau_\chi=1/\Gamma_\chi$, where $\Gamma_\chi$ is given by
eq.
(\ref{gamp}). The main difference between this result and the contribution from
the bosonic decay channel comes from the fermion distribution functions. 

\begin{center}
\begin{figure}[ht]
\scalebox{1.0}{\includegraphics{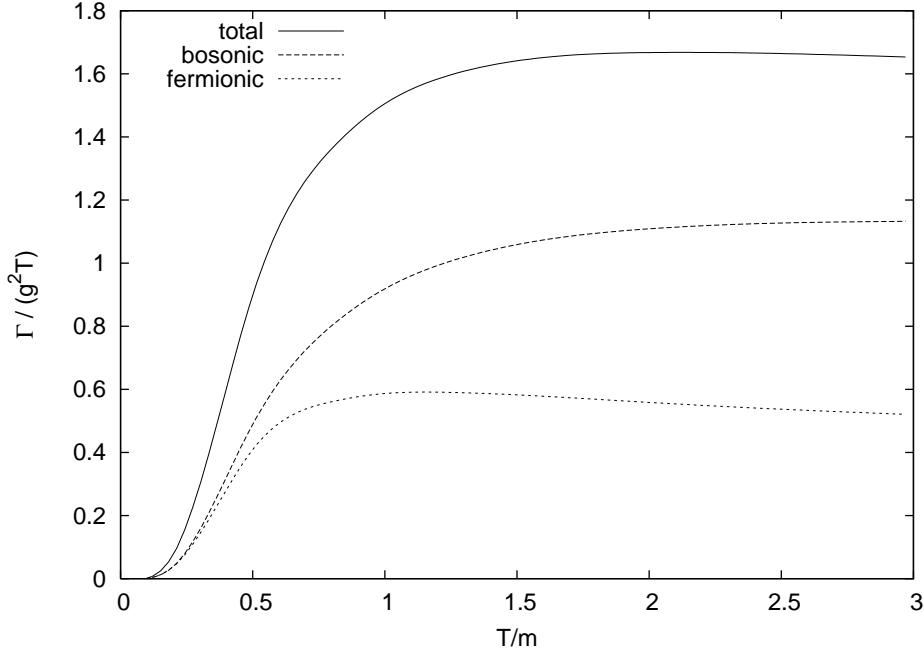}}
\caption{The bosonic and fermionic contributions to the dissipation coefficient 
plotted as a function of temperature for $h=0.79$.}
\label{figfplot}
\end{figure}
\end{center}

Figure \ref{figfplot} shows a comparison between the intermediate bosonic and
fermionic decays of the inflaton. The total dissipation coefficient from all
decays at high temperature is
\begin{equation}
\Gamma\approx 0.97\,{g^2\over h^2}\,T
\qquad h^{-1}m_\chi> T\gg m_\chi.
\end{equation}
When $T>h^{-1}m_\chi$, the two-loop contribution to the $\chi$ self-energy
corresponding to decays such as $\phi\to\eta yy$ becomes important. The size of
this contribution to the self-energy is known from previous work to be
$O(g^2h^2T^2)$ \cite{berera98}, leading to the dissipation coefficient of
Hosoya and Sakagami  $\Gamma=O(h^{-2}m_\chi^2 T^{-1})$ \cite{hosoya84}.

In the low temperature limit the integrals are dominated by small $k$ and
$\omega$. We have $\tilde\rho_\chi\approx 2\tilde\alpha/m_{\tilde\chi}$,
and
\begin{equation}
\Gamma(\phi\to y\tilde y)\approx
{g^2T^3\over m_{\tilde\chi}^4}
\int{d^3 k\over (2\pi)^3}\int_0^\infty {d\omega\over 2\pi}
8\,\beta^4{\rm tr}(\tilde\alpha^2)e^{-\beta\omega}
\end{equation}
Using the expression for $\alpha$ in (\ref{alzero}) and (\ref{alone}), 
\begin{equation}
\Gamma(\phi\to y\tilde y)\propto
\,g^2h^4\,{T^5\over m_{\tilde\chi}^4}
\end{equation}
The bosonic decay rate, which is $\propto T^3$, dominates at low temperatures.

\subsection{Dissipation with the exponential decay propagator}

The results have so far assumed that the heavy field has a thermal propagator.
We consider now what happens if we use the exponential decay approximation
for the propagator to calculated the friction coefficient $\Gamma_{MS}$ which
was discussed in section  IIIC. Although the heavy field is now not in
equilibrium, the light
field thermalises on a different timescale which we suppose is sufficiently
short to keep the light field in equilibrium.

For the bosonic decay we obtain the contribution  $\Gamma_{MS}(\phi\to2y)$ to
the dissipation coefficient from the decay $\phi\to\chi\to yy$ by using eq
(\ref{gammams}),
\begin{equation}
\Gamma_{MS}(\phi\to2y)=16 g^4\phi^2\int {d^3k\over (2\pi)^3}
{\alpha_y\over \omega_k^6},
\end{equation}
The self-energy $\alpha_y$ is determined by the results of
section IV. The dissipation coefficient can be obtained analytically,
\begin{equation}
\Gamma_{MS}(\phi\to2y)=
{h^2g^2\over 16\pi^2}\left({g^4\phi^4\over m_\chi^3}\right)
\left(1+\left({4T\over m_\chi}\right)^5 f\left({2T\over m_\chi}\right)\right)
\end{equation}
where
\begin{equation}
f(a)={2\over \pi}\int_0^\infty {x^4\over (1+a^2x^2)^4}{1\over e^x-1}.
\end{equation}
Similarly, the contribution  $\Gamma_{MS}(\phi\to2\tilde y)$ to
the dissipation coefficient from $\phi\to\chi\to \tilde y\tilde y$ becomes
\begin{equation}
\Gamma_{MS}(\phi\to2\tilde y)=
{h^2g^2\over 16\pi^2}\left({g^2\phi^2\over m_\chi}\right)
\left(1+\left({4T\over m_\chi}\right)^5 \tilde f\left({2T\over
m_\chi}\right)\right)
\end{equation}
where
\begin{equation}
\tilde f(a)={2\over \pi}\int_0^\infty {x^4\over (1+a^2x^2)^4}{1\over e^x+1}.
\end{equation}
The main application of these results would be at low temperature and the
temperature corrections are relatively small.

For the fermionic decay $\phi\to\tilde\chi\to y\tilde y$, we can make use of
the symmetry between the decay
width of the fermion $\tilde\chi$ given by eq. (\ref{gamp}) and the decay width
of the boson $\chi$ when $m_\chi=m_{\tilde\chi}$. We obtain the contribution
to the dissipation coefficient,
\begin{equation}
\Gamma_{MS}(\phi\to y\tilde y)=
{h^2g^2\over 8\pi^2}\,m_{\tilde\chi}
\left(1+\left({2T\over m_{\tilde\chi}}\right)^5 f\left({T\over
m_{\tilde\chi}}\right)  \right)
\end{equation}
In the zero-temperature limit, the total dissipation coefficient is then
\begin{equation}
\Gamma_{MS}=
{h^2g^2\over 16\pi^2}
\left(2+{g\phi\over m_\chi}+
{g^3\phi^3\over m_\chi^3}\right)\,g\phi.
\end{equation}
This is similar to the results obtained for non-supersymmetric models by 
Morikawa and Sasaki \cite{Morikawa:1984dz} and by Berera
and Ramos \cite{Berera:2001gs,berera02}, although the coefficient calculated
here is significantly larger due to the large number of decay channels
available to inflaton in the supersymmetric theory. In addition, there are
small numerical differences between different results due to slightly
different prefactors in the exponential decay approximation.

\section{Entropy production during inflation}

The main application for our results is to the inflationary universe. In this
section we shall make some general remarks and check the consistency of some
of the assumptions that we have made. During the slow evolution which is
characteristic of inflation, we replace the inflaton equation by
\begin{equation}
3H(1+r)\dot\phi+V'=0
\end{equation}
We have parameterised the effectiveness of dissipation by taking the ratio of
the dissipation coefficient to the expansion rate,
\begin{equation}
r={\Gamma\over 3H}.
\end{equation}
The energy density during inflation is dominated by the inflaton potential
energy $V$, and the expansion rate is given by
\begin{equation}
3H^2={8\pi\over m_p^2}V
\end{equation}
By combining these equations, we can compare $\dot\phi$ to $H$,
\begin{equation}
{\dot\phi^2\over H^2}={\epsilon\,m_p^2\over 4\pi(1+r)^2}\label{phidot}
\end{equation}
where $\epsilon$ measures the slope of the potential,
\begin{equation}
\epsilon={m_p^2\over 16\pi}\left({V'\over V}\right)^2.
\end{equation}
The second time derivative of the scale factor is positive for $\epsilon<1+r$.

The temperature of the radiation field can be determined from energy momentum
conservation. The energy lost by the inflaton is converted into heat and
increases the entropy density of the radiation $s$ \cite{Hall:2003zp},
\begin{equation}
T\dot s+3H Ts=\Gamma\dot\phi^2\label{entropy}
\end{equation}
The radiation is a relativistic gas, with
\begin{equation}
s=\frac43aT^3
\end{equation}
where $a$ is the effective radiation constant, taking into account the number
of degrees of freedom. We see immediately that for $\Gamma\propto T^3$, the
temperature does not fall to zero, but reaches a lower limit $T_r$,
\begin{equation}
T_rs=r\dot\phi^2.\label{trs}
\end{equation}
If $\Gamma$ falls off faster than $T^4$, then the cosmological redshift would
drive the temperature downwards and the radiation into the vacuum state. 

We shall consider this low temperature regime now in more detail, using the
thermal expression for the dissipation coefficient,
\begin{equation}
\Gamma=\gamma g^2h^4{T^3\over m_\chi^2}.
\end{equation}
The temperature falls to the value $T_r$ given by eq. (\ref{trs}). When we
eliminate $\dot\phi$ using (\ref{phidot}),
\begin{equation}
{T_r\over H}={\gamma g^2 h^4\over 16\pi a}
{\epsilon\over (1+r)^2}{m_p^2\over m_\chi^2}
\end{equation}
Further information can be obtained from the amplitude of fluctuations in the
cosmic microwave background. These have their origin as thermal fluctuations
during inflation if $T>H$. Theory gives an amplitude $\Delta$ for a given mode,
where
\begin{equation}
\Delta^2=c{H^2T(H\Gamma)^{1/2}\over\dot\phi^2},
\end{equation}
evaluated at the time that the mode crosses the horizon, and $c\approx 0.045$.
The measured amplitude $\Delta\approx5.4\times 10^{-5}$ on a $500\,\hbox{MPc}$
scale relevant to the large scale structure. We can use $\Delta$ to eliminate
the expansion rate and obtain
\begin{equation}
{T_r\over m_p}=\left({1\over 256\pi a c^2}\right)^{1/6}
{\epsilon^{1/2}\over 1+r}\,\Delta^{2/3}
\end{equation}

The range of temperatures for which the dissipation coefficient calculation can
be
applied and still be consistent with the low temperature approximation is
$H<T<m_\chi$. This translates to limits on the values of $m_\chi$ which are
consisten when
\begin{equation}
g h^2>\left({4\pi a\over c}\right)^{1/3}\gamma^{-1/2}\Delta^{2/3}
\approx 5.1\times 10^{-2}.
\end{equation}
It is quite possible for supersymmetric models of inflation to have coupling
constants in this range.

\section{conclusions}

We have taken a detailed look at the linear dissipation term in the equation of
motion for an inflaton field embedded in a supersymmetric model. The
dissipation was caused by the decay of the inflaton into light particles
during inflation with a heavy field $\chi$ acting as intermediary.
\begin{itemize}
\item For temperature $T>m_\chi$, our results are consistent with a wide body
of previous work. Some differences in the temperature dependence can be
attributed to the special features of the interactions which play an important
role for temperatures close to the heavy particle mass.
\item For $H<T<m_\chi$, where $H$ is the expansion rate, we find significant
differences between the dissipation coefficient in the thermal background and
the dissipation coefficient obtained assuming exponential real-time decay of
the $\chi$ propagator
\cite{Morikawa:1984dz,berera02,Berera:2004kc}. 
In particular, the dissipation coefficient in the thermal background does
not approach a constant as $T\to 0$ and $H\to 0$, but instead behaves as a
power law $\propto T^3$.
\end{itemize}
Thermal propagators decay exponentially for only a limited time before
switching over to a power law decay. The switch over occurs at a logarithmic
factor times the exponential decay time. A purely exponential decay might
possibly represent an approximation to the non-equilibrium behaviour of the
system if the dynamical timescale of the inflaton where shorter than the
thermalisation timescale of the heavy field. We have therefore given results
for both the thermal and exponential propagators.

We have argued that the low temperature dissipation coefficient obtained using
the thermal propagator can still have an important effect on inflationary
models. For a significant range of parameters, the dissipation heats the
radiation to
the $H<T<m_\chi$ range. Thermal fluctuations are then more important than
vacuum fluctuations and become the source of primordial density perturbations
 \cite{moss85,bererafang95,berera95,berera96,berera00,taylor00,Hall:2003zp}. 

Dissipation is far stronger for the exponential propagator and warm inflation
is realised for a far larger range of parameters. This case has been studied
in previous work which has divided up the parameter ranges into those leading
to cold and warm inflation
\cite{Bastero-Gil:2004tg,Hall:2004zr,Bastero-Gil:2005gy}. There remains much
to understand about the non-equilibrium physics of reheating during the
inflationary era, but the results which we have obtained so far indicate that
dissipation is important during inflation for supersymmetric models with
moderate values of the coupling constants.

\acknowledgments

We would like to acknowledge many helpful discussions
with Arjun Berera, Rudnei Ramos and Lisa Hall.

\appendix

\section{Properties of thermal propagators}

This appendix lists some important properties of the thermal equilibrium
propagators. We begin with some of the most basic rules,
\begin{itemize}
\item Kubo-Martin-Schwinger periodicity relations,
\begin{eqnarray}
\Pi_{21}&=&e^{\beta\omega}\Pi_{12}\label{per}\\
\Sigma_{21}&=&-e^{\beta\omega}\Sigma_{12}
\end{eqnarray}
\item The Kobes-Semenoff rules,
\begin{eqnarray}
{\rm Im}(\Pi_{11})&=&-\frac{i}2\left(\Pi_{12}+\Pi_{21}\right)\label{ks}\\
{\rm Im}(\Sigma_{11})&=&-\frac{i}2\left(\Sigma_{12}+\Sigma_{21}\right)
\end{eqnarray}
\end{itemize}
The Kobes-Semenoff rules follow from the maximum time rule. 

The periodicity relations imply that the off-diagonal components of the
self-energies can be expressed in terms of functions $\alpha(P)$
and $\tilde\alpha(P)$,
\begin{eqnarray}
\Pi_{21}=-i(1+n)\alpha&&\Pi_{12}=-in\alpha\label{pa}\\
\Sigma_{21}=-i(1-\tilde n)\tilde\alpha&&\Sigma_{12}=i\tilde n\tilde\alpha
\label{sa}
\end{eqnarray}
where
\begin{equation}
n(\omega)={1\over e^{\beta\omega}-1},
\qquad \tilde n(\omega)={1\over e^{\beta\omega}+1}
\end{equation}
are occupation numbers for bosons and fermions. The off-diagonal components of
the full two-point functions can also be expressed in terms of a real
function, the spectral density $\rho$,
\begin{eqnarray}
G_{21}=i(1+n)\rho&&G_{12}=in\rho\label{bspec}\\
S_{21}=i(1-\tilde n)\tilde\rho&&S_{12}=-i\tilde n\tilde\rho.\label{fspec}
\end{eqnarray}
If we discard the real part of the vacuum polarisation, then the 
Kobes-Semenoff rules imply that the spectral density $\rho$
and the function $\alpha$ are related by
\begin{equation}
\rho=i(P^2+m^2+i\alpha)^{-1}-i(P^2+m^2-i\alpha)^{-1}\label{specb}
\end{equation}
The spectral density typically has a rich variety of poles and branch cuts. For
a pole at $\omega=\pm\omega_p\pm i\tau^{-1}$, we can regard $\omega_p$ as the
energy of the particle state and $\tau$ as the relaxation time. In
perturbation theory, 
$\omega_p=(p^2+m^2)^{1/2}$ and 
$\tau=2\omega_p\,\alpha(p,\omega_p)^{-1}$.

For fermions, the spectral density
\begin{equation}
\tilde\rho=i(\gamma^\mu P_\mu+m+i\tilde\alpha)^{-1}
-i(\gamma^\mu P_\mu+m-i\tilde\alpha)^{-1}.\label{specf}
\end{equation}
We can use rotational invariance to express $\tilde\alpha$ in
terms of three scalar components,
\begin{equation}
\tilde\alpha=\tilde\alpha_0\sigma_0
+\tilde\alpha_1\sigma_1+\tilde\alpha_m m
\end{equation}
where the gamma-matrices have been adapted to a momentum frame,
\begin{eqnarray}
\sigma_0&=&\gamma^0\omega-\gamma\cdot\hat{\bf p}\,p\label{sig}\\
\sigma_1&=&\gamma\cdot\hat{\bf p}\,\omega-\gamma^0p
\end{eqnarray}
These combinations satisfy $-\sigma_0^2=\sigma_1^2=P^2$ and 
$\{\sigma_0,\sigma_1\}=0$. The poles of the spectral function are again given
approximately by 
$\pm\omega_p\pm i\tau^{-1}$, but now
\begin{equation}
\omega_p\tau^{-1}=(\tilde\alpha_0+\tilde\alpha_m)m^2\label{fgam}
\end{equation}
to leading order in $\tilde\alpha$. The relaxation time does not depend on
$\tilde\alpha_1$.

Spectral densities for a free theory can easily be obtained by taking  the
limit $\alpha\to 0$,
\begin{eqnarray}
\rho&=&2\pi\,{\rm sgn}(\omega)
\,\delta(P^2+m^2)\label{freespecb}\\
\tilde\rho&=&2\pi\,{\rm sgn}(\omega)(-\gamma^\mu P_\mu+m)
\,\delta(P^2+m^2).\label{freespecf}
\end{eqnarray}
The limit depends on the sign of $\omega$ because
$\alpha(p,-\omega)=-\alpha(p,\omega)$.
\bibliography{paper.bib,damping.bib}

\end{document}